\documentclass[twocolumn,showpacs,preprintnumbers,amsmath,amssymb, prl, reprint, longbibliography]{revtex4-1}
\usepackage{amsmath}
\usepackage{stmaryrd}
\usepackage{txfonts}
\usepackage{amssymb}
\usepackage{mathrsfs}
\usepackage{graphicx}
\usepackage{dcolumn}
\usepackage{bm}
\usepackage{epsfig}
\usepackage{color}
\usepackage[colorlinks=true, linkcolor=blue, citecolor=blue, urlcolor=blue]{hyperref} 
\begin{document}
\preprint{\href{http://dx.doi.org/10.1103/PhysRevB.94.020402}{S. -Z. Lin, Phys. Rev. B {\bf 94}, 020402(R) (2016).}}

\title{Edge instability in a chiral stripe domain under an electric current and skyrmion generation}
\author{Shi-Zeng Lin}
\email{szl@lanl.gov}
\affiliation{Theoretical Division, Los Alamos National Laboratory, Los Alamos, New Mexico 87545, USA}

\begin{abstract}
Motivated by the recent experimental observations on the skyrmion creation by cutting chiral stripe domains under a current drive [Jiang {\emph{et al.}}, Science {\bf{349}}, 283 (2015)], we study the mechanism of skyrmion generation by simulating the dynamics of stripe domains. Our theory for skyrmion generation is based on the fact that there are two half skyrmions attached to the ends of a stripe domain. These half skyrmions move due to the coupling between the skyrmion topological charge and current. As a consequence, the stripe domain is bent or stretched depending on the direction of motion of the half skyrmions. For a large current, skyrmions are created by chopping the stripe domains via strong bending or stretching. Our theory provides an explanation to the experiments and is supported by the new experiments. Furthermore, we predict that skyrmions can also be generated using a Bloch stripe domain under a spin transfer torque which can be realized in B20 compounds.
\end{abstract}
\pacs{75.10.Hk, 75.25.-j, 72.25.-b, 75.78.-n}
\date{\today}
\maketitle

A skyrmion in magnets is a stable spin texture which behaves like a particle at mesoscale \cite{Bogdanov89,Bogdanov94}. The skyrmion lattice has been observed in several classes of compounds without inversion symmetry including metals \cite{Muhlbauer2009,Yu2010a}, semiconductors \cite{Yu2011}, and insulators \cite{Adams2012,Seki2012}. Skyrmions can also be stabilized in multilayer systems where the inversion symmetry is broken at the interface \cite{Heinze2011,Chen2015,Jiang17072015}. Skyrmions respond to various external stimuli, such as electric current/field \cite{Jonietz2010,Yu2012,Schulz2012,White2012,PhysRevLett.113.107203}, thermal gradient \cite{Kong2013,Lin2014PRL,Mochizuki2014}, magnetic field, etc. The observations that skyrmions can also be driven by a low electric current density \cite{Jonietz2010,Yu2012,Schulz2012} make skyrmions prime candidates for the next generation spintronic devices. ~\cite{Fert2013,nagaosa_topological_2013}

For spintronic applications, it is required to generate skyrmions in a controlled way by an electric current. Several theoretical proposals have been put forward in the past few years \cite{szlin13skyrmion1,zhou_reversible_2014,sampaio_nucleation_2013,iwasaki_current-induced_2013}. On the experimental side, it was demonstrated that skyrmions can be nucleated or removed with local spin-polarized currents from a scanning tunneling microscope \cite{Romming2013}. Recently it was shown beautifully that skyrmions can be created by passing stripe domains stabilized in heavy metal/ultrathin ferromagnet/insulator trilayers through a geometrical constriction under a current pulse. \cite{Jiang17072015} The stripe domains at the constriction are dissected into skyrmions. It was argued that the inhomogenous current transverse to the stripe domain generated by the constriction is responsible for the skyrmion generation. The detailed mechanism of skyrmion generation however is not clear, because the magnetization dynamics is within nanosecond time scale which is beyond the experimental resolution.

In this work, we proposal an alternative explanation to the experiments. \cite{Jiang17072015} Our explanation relies on the observation that there are two half skyrmions at the ends of a stripe domain. These half skyrmions move under currents, therefore the stripe is bent. For a high current, the stripe is cut into skyrmions. In our theory, the geometry constriction is helpful to attain a large current density, but is not essential for skyrmion generation. This is consistent with the latest experiments with constriction width much bigger than the stripe domain width. \cite{jiang_mobile_2016} Moreover the predicted transverse motion of skyrmions to the current, which is essential for the skyrmion generation, has also been observed in the new experiments. \cite{jiang_direct_2016}

\begin{figure}[b]
\psfig{figure=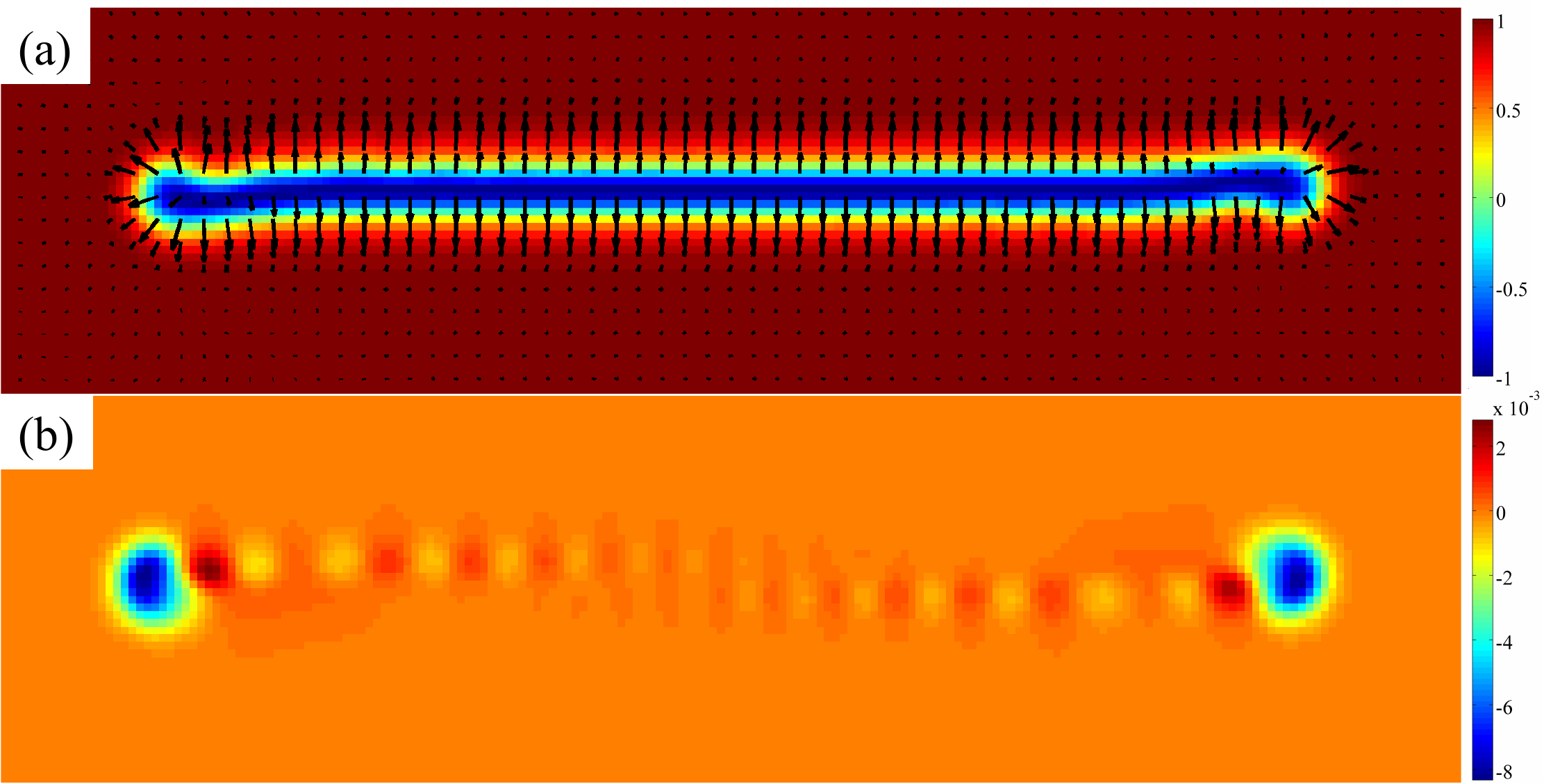,width=\columnwidth}
\caption{(color online) (a) A typical N\'{e}el chiral stripe domain in the ferromagnetic background, and (b) the corresponding topological charge density $q(\mathbf{r})$. There are two half skyrmions at the ends of the stripe. There is also a small $q(\mathbf{r})$ in the middle of the stripe because the stripe is not straight. In (a) color represents $n_z$ and the vector field denotes the $n_x$ and $n_y$ components.
} \label{f1}
\end{figure}

 \begin{figure*}[t]
\psfig{figure=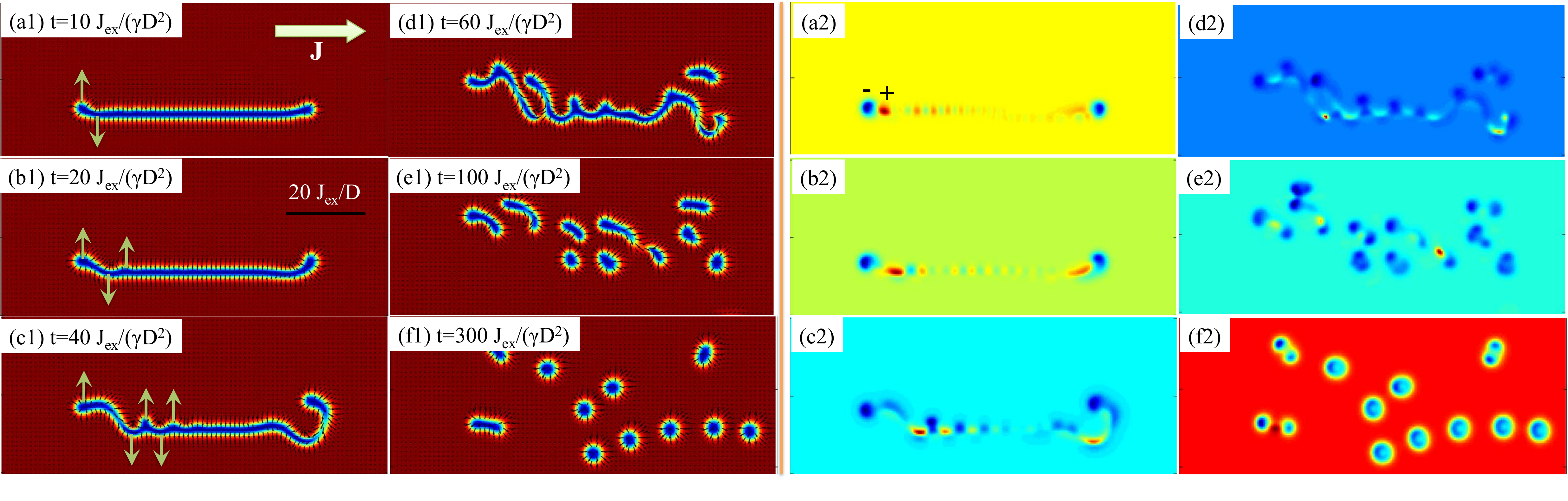,width=18.5cm}
\caption{(color online) Left: dynamics on the bending of a N\'{e}el stripe domain. The spin Hall torque produces a zigzag distortion of the stripe domain (a1 - d1) and eventually the stripe domain is converted into skyrmions (e1 - f1). The green arrows denote the direction of local velocity. The dark line denotes length scale in the plots.  Right: the corresponding skyrmion topological charge density $q(\mathbf{r})$. Note that $q(\mathbf{r})$ is negative (positive) in the blue (red) region. Here $q(r)=0$ in the ferromagnetic region and the background color changes because a different scale is used.  Here $H_a=0.6 D^2/J_{\mathrm{ex}}$ and $J=0.2 D e/\hbar$. See Ref. \onlinecite{supplementS2S} for a movie.
} \label{f2}
\end{figure*}

The magnetism in heavy metal/ultrathin ferromagnet/insulator trilayers can be described by the following Hamiltonian density in two dimensions \cite{thiaville_dynamics_2012,PhysRevB.90.184427}
\begin{equation}\label{eq1}
\mathcal{H}=\frac{J_{\mathrm{ex}}}{2}\left(\nabla\mathbf{n}\right)^2+D[n_z(\nabla\cdot \mathbf{n})-(\mathbf{n}\cdot\nabla)n_z]-\mathbf{H}_a\cdot\mathbf{n},
\end{equation}
where $J_{\mathrm{ex}}$ is the exchange interaction, $D$ is the Dzyaloshinskii-Moriya interaction (DMI) due to the breaking of the inversion symmetry at the interface and $\mathbf{H}_a=H_a\hat{z}$, with $\hat{z}$ a unit vector perpendicular to the film, is the external magnetic field. Here $\mathbf{n}$ is a unit vector representing the spin direction.  The DMI stabilizes N\'{e}el stripe domains (cycloidal rotation of spins), see Fig. \ref{f1}. At the ends of the domain, the spins are noncoplanar and the skyrmion topological charge density $q(\mathbf{r})=\mathbf{n}\cdot(\partial_x\mathbf{n}\times\partial_y\mathbf{n})/(4\pi)$ is nonzero. The total skyrmion charge at one end is $-1/2$ and therefore the stripe domain can be regarded as an elongated skyrmion \cite{Ezawa2011}, i.e. a skyrmion is split into two halves and they are connected by the stripe domain.

The dynamics of spins in the presence of a spin Hall torque generated by the spin Hall effect is described by \cite{emori_current-driven_2013,ryu_chiral_2013,PhysRevB.90.184427,liu_spin-torque_2012}
\begin{equation}\label{eq2}
\partial_t\mathbf{n} =-\gamma  \mathbf{n}\times \mathbf{H}_{\mathrm{eff}}+\alpha  \mathbf{n}\times \partial_t\mathbf{n} +\frac{\hbar \gamma \theta_{\mathrm{sh}}}{2e d}\mathbf{n}\times \left[\mathbf{n}\times \left(\hat{z}\times\mathbf{J}\right)\right],
\end{equation}
where $\mathrm{H}_{\mathrm{eff}}\equiv-\delta\mathcal{H}/\delta \mathbf{n}$ is an effective field, $\mathbf{J}$ is the electric current, $\alpha$ is the Gilbert damping and $\gamma$ is the  gyromagnetic ratio. Here $\theta_{\mathrm{sh}}$ is the spin Hall angle and $d$ is the film thickness. We use $\alpha=0.1$ in simulations. To understand qualitatively the motion of a stripe domain under a spin Hall torque,  we derive the equation of motion for a rigid half skyrmion by using the Thiele's approach \cite{Thiele72}
\begin{equation}\label{eq3}
Q \hat{z}\times \mathbf{v}+\alpha  \eta  \mathbf{v}=\frac{\hbar \gamma \theta_{\mathrm{sh}}}{2e d} J \mathbf{Y},
\end{equation}
where $\mathbf{v}$ is the velocity of the half skyrmion and $\eta =\int' \left(\partial_\mu \mathbf{n}\right)^2  dr^2/(4\pi)$ and $Y_\mu=\left(\hat{z}\times \hat{\mathbf{J}} \right)\cdot \int'\left(\partial_\mu \mathbf{n}\times\mathbf{n} \right)d r^2/(4\pi)$ with $\mu=x,\ y$ and the integrated topological charge $Q\equiv\int' q(\mathbf{r})dr^2=-1/2$. The integration is restricted to the region around the half skyrmion. $\hat{\mathbf{J}}$ is the unit vector along the current direction. We decompose $\mathbf{v}$ and $\mathbf{Y}$ into the components parallel and perpendicular to the current. For a N\'{e}el stripe domain, $Y_\perp=0$.  The velocity of the half skyrmion is given by $v_\perp=-Q Y_\parallel \hbar \gamma \theta_{\mathrm{sh}} J/[2ed(Q^2+(\alpha\eta)^2)]$ and $v_\parallel=\alpha \eta Y_\parallel \hbar \gamma \theta_{\mathrm{sh}} J/[2ed(Q^2+(\alpha\eta)^2)]$. The half skyrmion has velocity component perpendicular to the current direction. Equation \eqref{eq3} describes only qualitatively the dynamics of the half skyrmion, because both $\mathbf{Y}$ and $\eta$ depend on the shape of the half skyrmion, which changes under a current drive.

The full dynamics of a stripe domain subjected to a current along the direction of the stripe is obtained by numerical simulations of Eq. \eqref{eq2}. \cite{supplementS2S} The dynamical evolution of the stripe domain is displayed in Fig. \ref{f2}. According to Eq. \eqref{eq3}, the half skyrmions at both ends move up, which causes the stripe domain to bend upward [see Fig. \ref{f2} (a1)]. The spin in the bending region is noncoplanar and a non-quantized positive skyrmion charge is accumulated there [see Fig. \ref{f2} (a2)]. Such a positive skyrmion charge moves downward [see Fig. \ref{f2} (b1)]. Such downward bending of the stripe domain generates negative skyrmion charge and it moves upward [see Fig. \ref{f2} (b2)]. This process repeats and produces a zigzag distortion of the stripe domain. [see Fig. \ref{f2} (c1) and (d1)] For a weak current, such a zigzag distortion is balanced by the elastic energy of the stripe domain. However for a strong current, the zigzag distortion breaks the stripe into small domains. These domains relax and become skyrmions [see Fig. \ref{f2} (e1) and (f1)]. The threshold current to slice a stripe domain into skyrmions is about $J_t\approx 0.1\ D^2 e d/(J_{\mathrm{ex}}\hbar\theta_{\mathrm{sh}} )$ in the field region $0.5 D^2/J_{\mathrm{ex}}\le H_a\le 0.8 D^2/J_{\mathrm{ex}}$. The process is asymmetric with respect to the center of the stripe domain because of the nonzero line tension of the stripe domain, i.e. the stripe domain grows or shrinks in a clean system depending on the magnetic field as will be discussed below \cite{supplementS2S}.  

 \begin{figure*}[t]
\psfig{figure=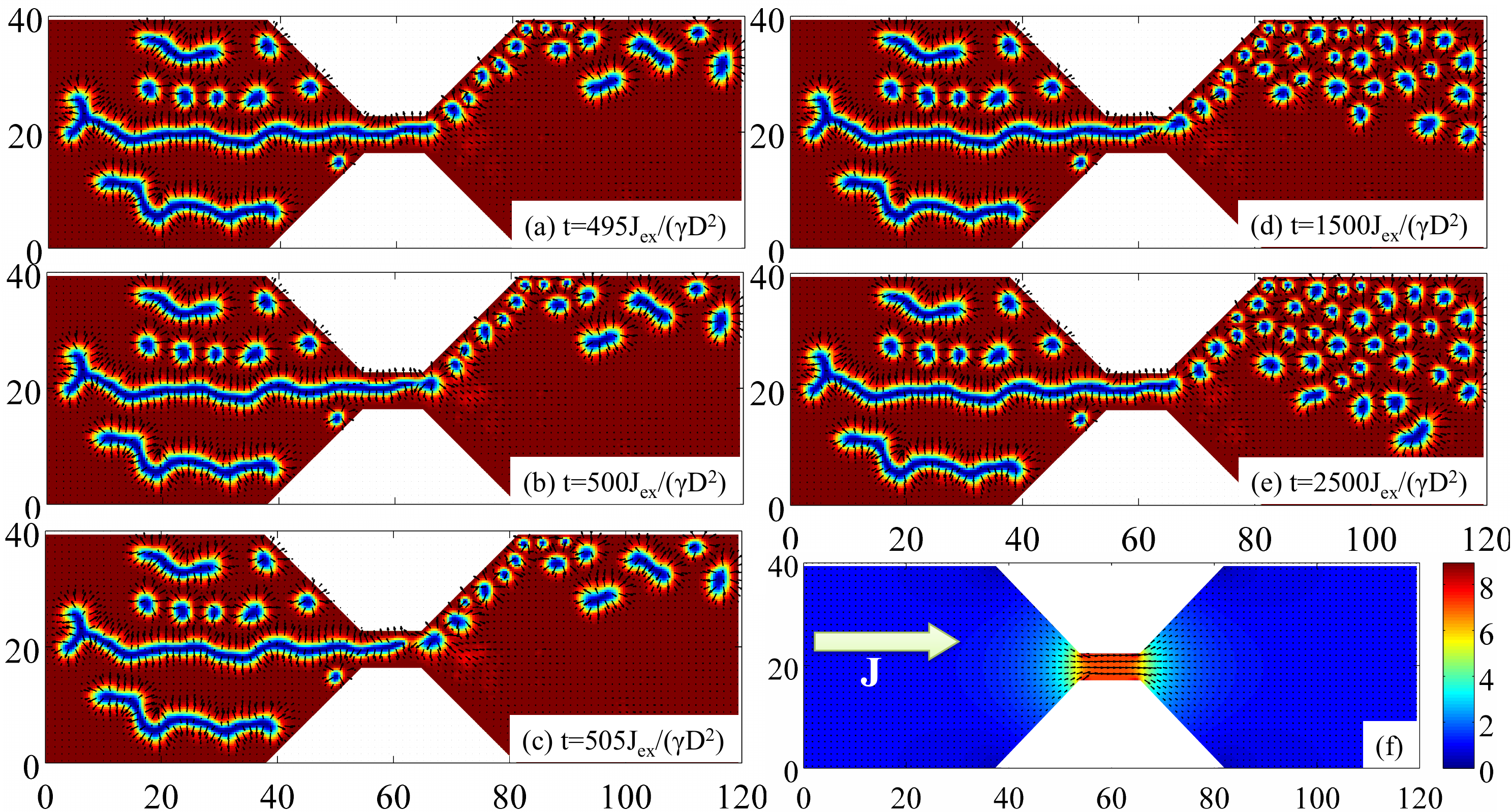,width=18.5cm}
\caption{(color online) (a - e) Generation of skyrmions by chopping N\'{e}el stripe domains with spin Hall torque in a sample with a geometrical constriction. We introduce disorders into the system by adding a random easy axis anisotropy term, $-AS_z^2/2$ with $\langle A(\mathbf{r})A(\mathbf{r}')\rangle=0.64\delta(\mathbf{r}-\mathbf{r}')D^4/J_{\mathrm{ex}}^2$, to the Hamiltonian in Eq. \eqref{eq1}. The skyrmions at the left side are created in the relaxation process and are pinned.  (f) The distribution of normalized current density inside the sample. Arrows denote the current direction and color represents its amplitude. The current density at the right/left edge is $J_x=0.04 \ D^2 e d/(J_{\mathrm{ex}}\hbar\theta_{\mathrm{sh}} )$ and $H_a=0.7D^2/J_{\mathrm{ex}}$. The length is in unit of $J_{\mathrm{ex}}/D$. See Ref. \onlinecite{supplementS2S} for a movie.
} \label{f3}
\end{figure*}

Our simulations show that a strong homogenous current can cut a N\'{e}el stripe domain into skyrmions. To achieve a strong current, we consider a geometrical constriction. The current distribution is governed by $\nabla^2V=0$ subjected to the boundary condition $V=V^+$ at the left edge and $V=0$ elsewhere. Here $V$ is the voltage and the electric current is $\mathbf{J}=-\sigma\nabla V$ with an electric conductivity $\sigma$. \cite{SZLinMonopole2016} The current intensity at the constriction is $w/w_c$ times stronger than that at the left/right side, with $w$ ($w_c$) being the width of the sample at left/right edges (constriction). The dynamical process of skyrmion generation by cutting the stripe domains is displayed in Fig.  \ref{f3}.  We prepare three stripe domains at the left side. The skyrmions at the left side are generated in the relaxation process and are pinned by defects. We then inject a dc current at the left side. Once one end of a stripe domain with a half skyrmion enters into the constriction, the half skyrmion is split off from the stripe. It then relaxes into a skyrmion and moves along the upper edge of the sample due to the driving force by current and the boundary potential. Meanwhile a new half skyrmion is generated at the end of the stripe domain in the process of spin relaxation. These processes repeat and we have continuous generation of skyrmions by chopping stripe domains at the constriction. The skyrmion generation is due to the bending of the stripe domain caused by the transverse motion of the half skyrmion. Inside the geometric constriction, the bending is balanced by the geometric confinement due to the boundary. Therefore the skyrmion generation only occurs at the entrance or exit of the constriction. In the presence of disorders, the main body of the stripe domains are strongly pinned. Away from the constriction where current is weak, the skyrmions are also pinned. These pinned skyrmions together with the disorders affect the motion of skyrmions.

In Ref. \onlinecite{Jiang17072015},  it is argued that the inhomogenous current perpendicular to the stripe domain produces an effective spin Hall force, which cuts the stripe domain and produces skyrmions moving along the current direction. For a current along the stripe domain direction, the transverse current component can be induced by a geometric constriction. In this picture,  the geometric constriction is essential for the skyrmion generation.  Our theory provides an alternative explanation to the experiments. \cite{Jiang17072015} The stripe domains are bent because of the transverse motion of the half skyrmions at their ends in the presence of a spin Hall torque. For a large current, the half skyrmions are cut from the stripe domains and relax into full skyrmions. The predicted transverse motion of skyrmions has been observed in recent experiments. \cite{jiang_direct_2016} In contrast to Ref. \onlinecite{Jiang17072015},  our interpretation on the generation of skyrmion from a N\'{e}el stripe domain does not require inhomogeneity in the current density, therefore the geometry constriction is not essential but is only helpful to achieve a strong current density. The threshold current grows linearly with the width of the constriction. The new experiments \cite{jiang_mobile_2016} with different constriction width support our theoretical explanation. When the constriction width is much larger than that of the stripe domain, the transverse current component at the stripe domain is small [see Fig. \ref{f3} (f)]. Therefore the observation of the skyrmion generation in a wide constriction is incompatible with the explanation in Ref. \onlinecite{Jiang17072015}.    

 \begin{figure*}[t]
\psfig{figure=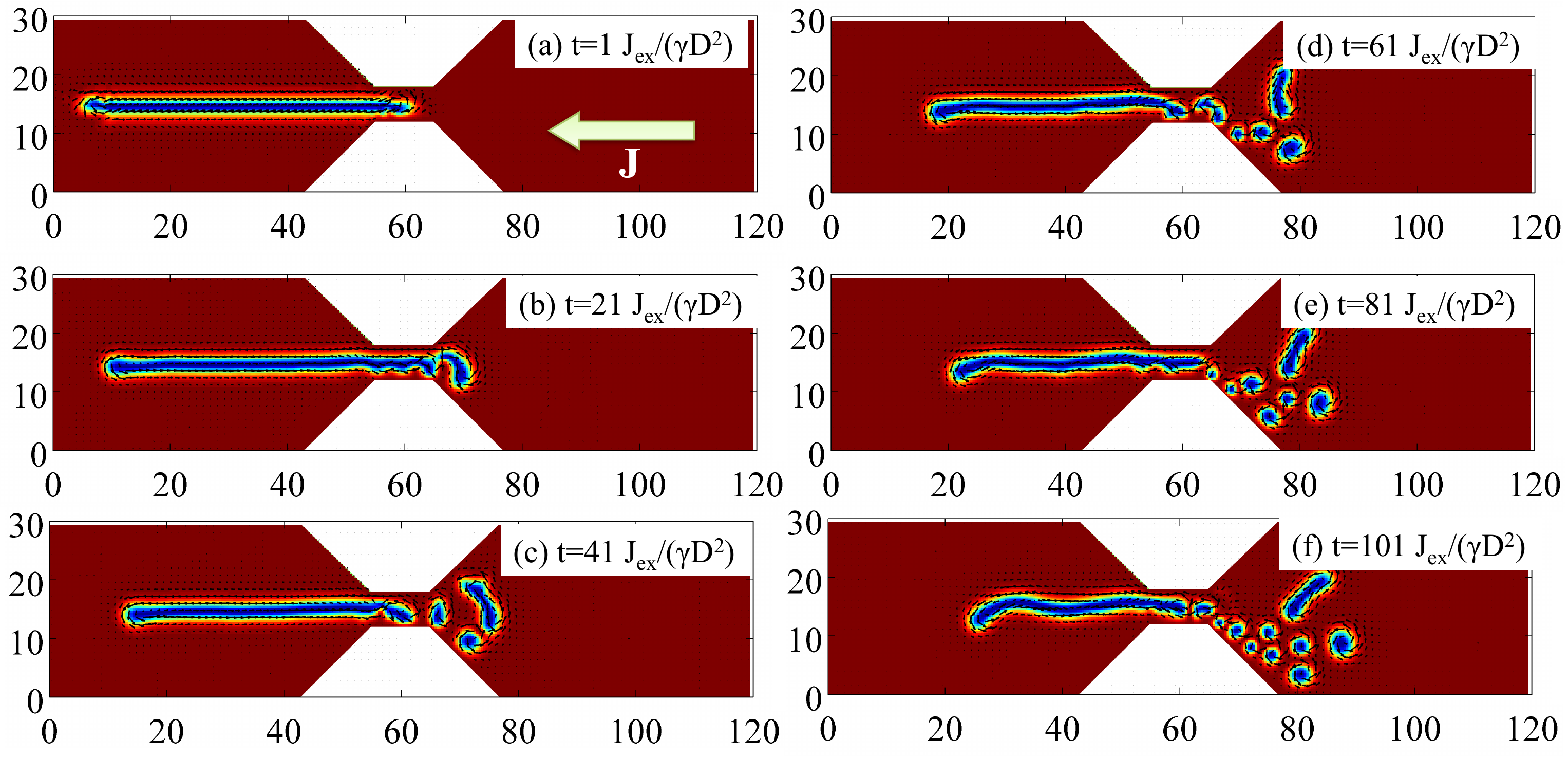,width=18.5cm}
\caption{(color online) Generation of skyrmions by stretching a Bloch stripe domain with an inhomogeneous spin transfer torque in a sample with a geometrical constriction. The current density at the right/left edge is $J_x=-0.4 D e/\hbar$ and $H_a=0.6 D^2/J_{\mathrm{ex}}$. See Ref. \onlinecite{supplementS2S} for a movie.
} \label{f4}
\end{figure*}

In experiments the skyrmions move almost parallel to the current \cite{Jiang17072015}, which is different from the simulation of a clean system in Fig. \ref{f1}. The longitudinal motion of skyrmions in experiments is due to the strong pinning potential created by defects and other pinned skyrmions. We perform simulations to study the effects of disorders. The defects are modeled by adding an uncorrelated random easy axis anisotropy $-AS_z^2/2$, with $\langle A(\mathbf{r})A(\mathbf{r}')\rangle=\kappa\delta(\mathbf{r}-\mathbf{r}')$, to the Hamiltonian in Eq. \eqref{eq1}. The disorders play following roles according to the simulations: (1) As shown in Ref. \onlinecite{supplementS2S} for $H_a>0.55 D^2/J_{\mathrm{ex}}$, the stripe domain shrinks into a full skyrmion by merging the two half skyrmions at the ends. While for $H_a<0.55 D^2/ J_{\mathrm{ex}}$, the stripe domain grows with time indicating a repulsion between two half skyrmions. For instance, even if we initialize the system with a single skyrmion in the ferromagnetic background, the skyrmion is split into two half skyrmions with separation increases with time. The threshold field $H_2=0.55 D^2/J_{\mathrm{ex}}$ is the field where the eigen-frequency of the internal mode with an angular momentum $l=2$ vanishes \cite{Bogdanov94,PhysRevB.90.094423, Lin_internal_2014}. Therefore the stripe domains and skyrmions cannot coexist in clean systems described by Eqs. \eqref{eq1}. The disorders pin the stripe domains, therefore stabilize the stripe domains. (2) Under a weak current, the stripe domain crawls and its length increases [see movie3.mp4 in Ref. \onlinecite{supplementS2S}]. Such a crawl motion of the stripe domain has also been observed in the experiments \cite{Jiang17072015}. For a strong current, the stripe domain breaks at the end because of the existence of the half skyrmion [see movie4.mp4 in Ref. \onlinecite{supplementS2S}]. This edge instability of the stripe domain is the same as that in a clean system. (3) The disorders provide an inhomogenous pinning force which may promote the skyrmion generation. For example, the main body of the stripe domain may be strongly pinned while the half skyrmion at its end is less pinned. It is easier to split the half skyrmion by a current with the main body of the stripe domain being fixed [see Fig. \ref{f3} and the corresponding movie]. (4) The disorders and the pinned skyrmions strongly affect the motion of skyrmions [see Fig. \ref{f3}, movie4.mp4  and movie5.mp4 in Ref. \onlinecite{supplementS2S}]. We remark that the mechanism of the skyrmion generation by chopping the half skyrmion at the end of the stripe domain does not change in the presence of defects.

We then generalize the skyrmion generation from N\'{e}el stripe domains to Bloch domains (spiral rotation of spins), which is relevant for the B20 compounds, such as MnSi and FeGe. The Hamiltonian of the system is $\mathcal{H}=\frac{J_{\mathrm{ex}}}{2}\left(\nabla\mathbf{n}\right)^2+D\mathbf{n}\cdot(\nabla\times\mathbf{n})-\mathbf{H}_a\cdot\mathbf{n}$. The dynamics of spins is governed by $\partial_t\mathbf{n} =-\gamma  \mathbf{n}\times \mathbf{H}_{\mathrm{eff}}+\alpha  \mathbf{n}\times \partial_t\mathbf{n} +\frac{\hbar\gamma}{2e}(\mathbf{J}\cdot\nabla)\mathbf{n}$, where the last term $(\mathbf{J}\cdot\nabla)\mathbf{n}$ is the adiabatic Slonczewskii-like spin transfer torque.

For the same reason, there are two half skyrmions at the ends of a Bloch stripe domain. \cite{supplementS2S} The Thiele's equation of motion for a half skyrmion in this case is $Q \hat{z}\times \left(\mathbf{v}+ \frac{ \hbar \gamma }{2 e}\mathbf{J}\right)=-\alpha  \eta  \mathbf{v}$. The velocity is $v_\parallel=-\gamma\hbar Q^2 J/[2e(Q^2+\alpha^2)]$ and $v_\perp=-\gamma\hbar \alpha Q J/[2e(Q^2+\alpha^2)]$. The half skyrmion travels almost antiparallel to the current for $\alpha\ll 1$. In the presence of a uniform current parallel to the stripe domain, the half skyrmions at both ends of the stripe domain move antiparallel to the current and therefore the whole stripe domain travels antiparallel to the current. The bending of the stripe domain is due to the transverse motion of the half skyrmion, which is weak for $\alpha\ll1$. It would require an enormous current density to break the stripe domain. We propose to apply an inhomogeneous current density to cut the stripe due to the disparity of the driving force at different ends. 

We consider the dynamics of a stripe domain in a sample with a constriction shown in Fig. \ref{f4}. Once one end of the stripe enters into the constriction, the half skyrmion at the edge is pulled by spin transfer torque due to current. The force acting on the half skyrmion at the constriction is $w/w_c$ times bigger than that at the other end. As a consequence, the stripe domain is stretched. For a large current, the half skyrmion at the constriction is pulled apart from the stripe. Then skyrmions are generated continuously by cutting the stripe domain at the constriction. The threshold current for skyrmion generation in this case is an order of magnitude bigger than that for a N\'{e}el stripe domain, because bending is more efficient than stretching a stripe domain in order to split the half skyrmion.

To summarize, we have studied the skyrmion generation from a N\'{e}el stripe domain. There are half skyrmions attached to the ends of a stripe domain. In the presence of spin Hall torque, the half skyrmion gains a velocity component perpendicular to the current. As a result, the stripe domain is bent. For a large current, the half skyrmion is cut from the stripe domain and it relaxes into a full skyrmion. Our theory provides a natural explanation to the experiments \cite{Jiang17072015} and is supported by the latest experiments \cite{jiang_direct_2016,jiang_mobile_2016}.  Skyrmions can also be generated using a Bloch stripe domain under a spin transfer torque which can be realized in B20 compounds.

\begin{acknowledgments}
The authors are indebted to Axel Hoffmann and Wanjun Jiang for sharing the experimental data prior to publication and for the helpful discussions. Computer resources for numerical calculations were supported by the Institutional Computing Program at LANL. This work was funded by the LANL Directed Research and Development program.
\end{acknowledgments}

\bibliography{reference}

\newpage
\clearpage
\appendix

\section{Details of numerical simulations}
We use dimensionless units in simulations: length is in units of $J_{\rm{ex}}/D$; energy is in units of $J_{\rm{ex}}^2/D$; magnetic field is in units of $D^2/J_{\rm{ex}}$; time is in units of $J_{\rm{ex}}/(\gamma D^2)$; The system is discretized with a grid size $0.4$; a smaller grid size is also used to check the accuracy of the results. We use the open boundary condition. Equations (2) and (5) in the main text are solved by an explicit numerical scheme developed in Ref. \onlinecite{Serpico01}.

\section{Bloch chiral stripe domain}
\begin{figure}[t]
\psfig{figure=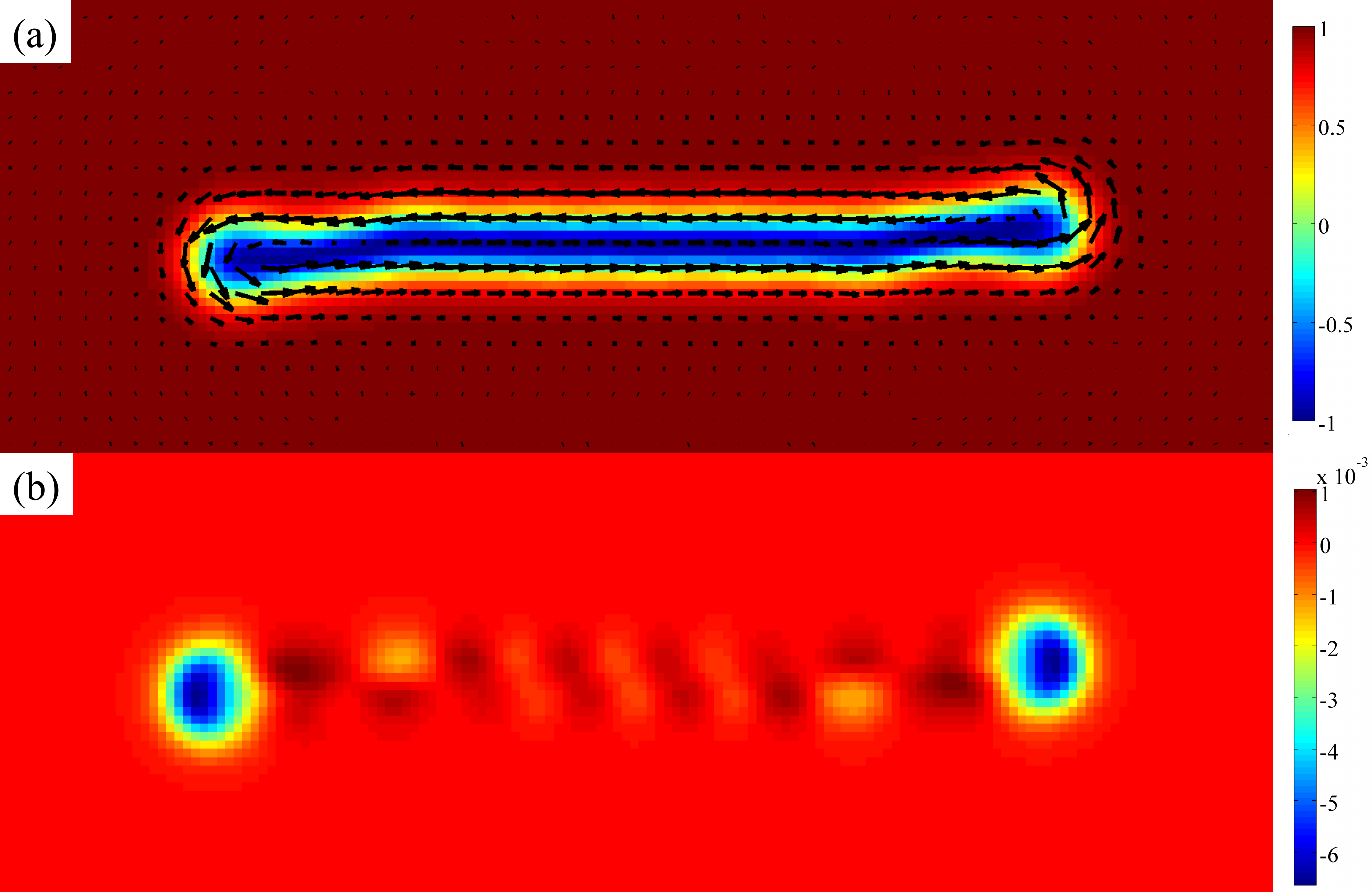,width=\columnwidth}
\caption{(color online) (a) A typical Bloch chiral stripe domain in the ferromagnetic background, and (b) the corresponding topological charge density $q(\mathbf{r})$. There are two half skyrmions at the ends of the stripe. There is also small $q(\mathbf{r})$ in the middle of the stripe because the stripe is not straight. In (a) color represents $n_z$ and the vector field denotes the $n_x$ and $n_y$ components.
} \label{fs0}
\end{figure}
A Bloch chiral stripe domain and the corresponding skyrmion topological charge density $q(\mathbf{r})$ for the Hamiltonian Eq. (4) in the main text is displayed in Fig. \ref{fs0}.

\section{Movie list}

 \begin{figure*}[t]
\psfig{figure=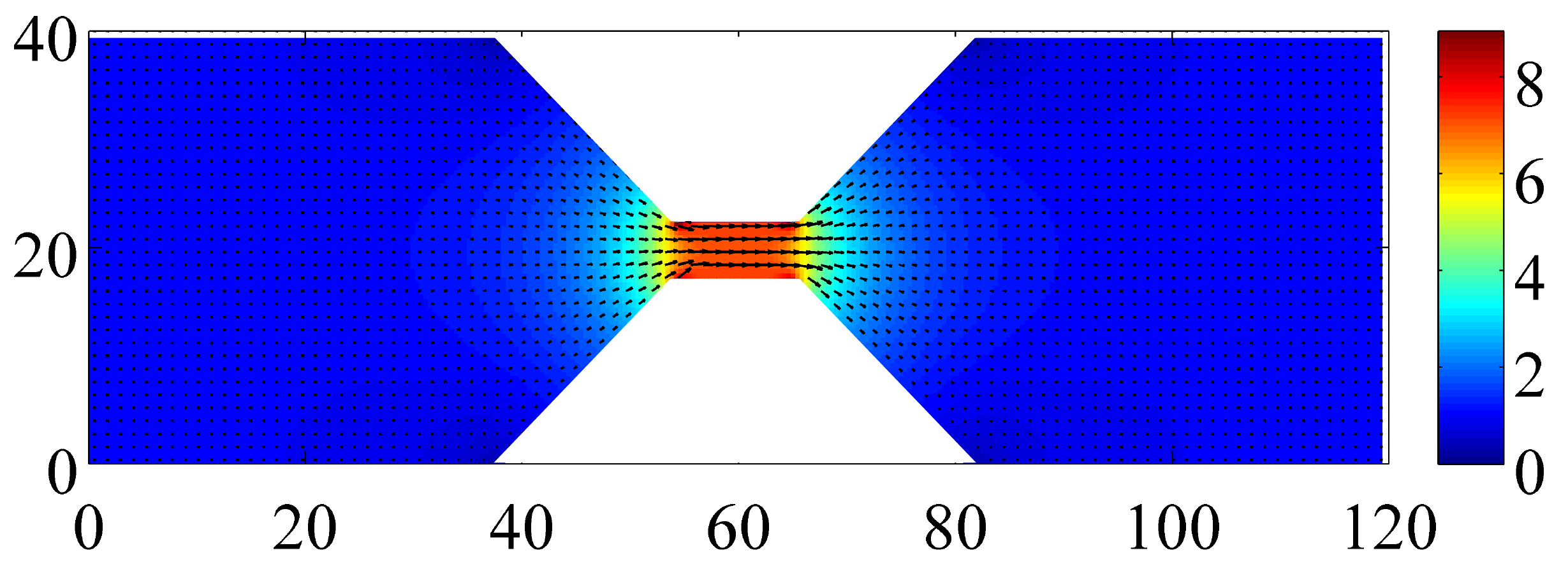,width=18.5cm}
\caption{(color online) Normalized current distribution inside the sample. Arrows denote the current direction and color represents its amplitude.
} \label{fs2}
\end{figure*}

Below we provide a list of movies on skyrmion generations from chiral stripe domains by an electric current.
\begin{description}
  \item[movie1.mp4] \hfill \\
  Stability of a chiral stripe domain in the ferromagnetic background when $H_a=0.8>H_2=0.55 D^2/J_{\mathrm{ex}}$. The stripe shrinks and rotates due to the Magnus force. Finally it contracts into a skyrmion. The total skyrmion charge does not change in this process. Here we have considered a Bloch stripe domain. The stability for a N\'{e}el domain is the same because the Bloch domain can be mapped to the N\'{e}el domain by global rotation of spin along the magnetic field axis by $\pi/2$.
  \item[movie2.mp4] \hfill \\
  Stability of a chiral stripe domain in the ferromagnetic background when $H_a=0.4<H_2=0.55 D^2/J_{\mathrm{ex}}$. We start from a skyrmion configuration and the skyrmion is split into a chiral stripe domain with two half skyrmions attached to its ends. Eventually the chiral stripe spans the whole system. 
  \item[movie3.mp4] \hfill \\
  Dynamics of a strip domain under a weak spin Hall torque. We introduce defects into the system by adding a random easy axis anisotropy term,  $-AS_z^2/2$ with $\langle A(\mathbf{r})A(\mathbf{r}')\rangle=0.64\delta(\mathbf{r}-\mathbf{r}')D^4/J_{\mathrm{ex}}^2$, to the Hamiltonian in Eq. (1) in the main text. The defects stabilize the stripe domain. We then inject a dc current into the system. The stripe domain starts to crawl. In certain region, the half skyrmion is split off the end the stripe domain and relaxes into a full skyrmion. In this simulation, we use a weak current $J=0.08 \ D^2 e d/(J_{\mathrm{ex}}\hbar\theta_{\mathrm{sh}} )$. The applied magnetic field is $H_a=0.7 D^2/J_{\mathrm{ex}}$ and we use the open boundary condition.

  \item[movie4.mp4] \hfill \\
 Dynamics of a strip domain under a strong spin Hall torque. We introduce disorders into the system by adding a random easy axis anisotropy term,  $-AS_z^2/2$ with $\langle A(\mathbf{r})A(\mathbf{r}')\rangle=0.64\delta(\mathbf{r}-\mathbf{r}')D^4/J_{\mathrm{ex}}^2$, to the Hamiltonian in Eq. (1) in the main text. The defects stabilize the stripe domain. We then inject a dc current into the system. The stripe domain is dissected into skyrmions. The dominant instability is at the ends of the stripe domain due to the existence of the half skyrmions there. The motion of skyrmion is affected by the defects. In this simulation, we use a weak current $J=0.12 \ D^2 e d/(J_{\mathrm{ex}}\hbar\theta_{\mathrm{sh}} )$. The applied magnetic field is $H_a=0.7 D^2/J_{\mathrm{ex}}$ and we use the open boundary condition.
  
  \item[movie5.mp4] \hfill \\
Movie of the skyrmion generation from N\'{e}el stripe domains under a spin Hall torque with a geometrical constriction corresponding to Fig. 3 in the manuscript. We introduce disorders into the system by adding a random easy axis anisotropy term,  $-AS_z^2/2$ with $\langle A(\mathbf{r})A(\mathbf{r}')\rangle=0.64\delta(\mathbf{r}-\mathbf{r}')D^4/J_{\mathrm{ex}}^2$, to the Hamiltonian in Eq. (1) in the main text. The skyrmions at the left side are created in the relaxation process and are pinned. The current distribution in the sample is displayed in Fig. \ref{fs2}. We have normalized the current density at edges to $1$. The current distribution can be scaled by a factor to achieve the required current density.

  \item[movie6.mp4] \hfill \\
Movie of skyrmion generation from a Bloch stripe domain with spin transfer torque corresponding to Fig. 4 in the manuscript. 
  
\item[movie7.mp4] \hfill \\
Movie of skyrmion generation from Bloch stripe domains under a spin transfer torque in a sample with a geometrical constriction. We introduce disorders into the system by adding a random easy axis anisotropy term,  $-AS_z^2/2$ with $\langle A(\mathbf{r})A(\mathbf{r}')\rangle=0.64\delta(\mathbf{r}-\mathbf{r}')D^4/J_{\mathrm{ex}}^2$, to the Hamiltonian in the main text.  The current density is $J= -0.15 D e/\hbar$ and the magnetic field is $H_a=0.7 D^2/J_{\mathrm{ex}}$.
\end{description}

\end{document}